\documentclass[pra,twocolumn,showpacs,preprintnumbers,amsmath,amssymb]{revtex4}

\usepackage{graphicx}
\usepackage{dcolumn}
\usepackage{bm}

\def \ed {\end{document}}
\def\Fbox#1{\vskip1ex\hbox to 8.5cm{\hfil\fboxsep0.3cm\fbox{%
\parbox{8.0cm}{#1}}\hfil}\vskip1ex\noindent}  
\def\be{\begin{equation}}\def\ee{\end{equation}}\def\bea{\begin{eqnarray}}\def\eea{\end{eqnarray}}\def\bse{\begin{subequations}}\def\ese{\end{subequations}}\newcommand{\BE}[1]{\begin{equation}\label{#1}}\newcommand{\BEA}[1]{\begin{eqnarray}\label{#1}}\newcommand{\BSE}[1]{\begin{subequations}\label{#1}}\let \= \equiv \let\*\cdot \let\~\widetilde \let\^\widehat \let\-\overline

\def\<{\left\langle}    \def\>{\right\rangle}\def\({\left(}          \def\){\right)} \def \[ {\left [} \def \] {\right ]} 

\begin{document}\preprint{APS/123-QED}\title{Spin Turbulence in a Trapped Spin-1 Spinor Bose--Einstein Condensate}\author{Kazuya Fujimoto $^1$ and Makoto Tsubota $^{1,2}$}\affiliation{$^1$Department of Physics, Osaka City University, 3-3-138 Sugimoto, Sumiyoshi-ku, Osaka 558-8585, Japan   \\$^2$ The OCU Advanced Research Institute for Natural Science and Technology (OCARINA),Osaka City University, 3-3-138 Sugimoto, Sumiyoshi-ku, Osaka 558-8585, Japan}\date{\today}

\begin{abstract}We numerically study spin turbulence in a two-dimensional trapped spin-1 spinor Bose--Einstein condensate, focusing on the energy spectrum. The spin turbulence in the trapped system is generated by instability of the helical structure of the spin density vector in the initial state. Our numerical calculation finds that in the trapped system the spectrum of the spin-dependent interaction energy in the ferromagnetic case exhibits a $-7/3$ power law, which was confirmed in a uniform system by our previous study. The relation between the $-7/3$ power law and the motion of the spin density vector is discussed by investigating the orbits of dynamical variables in the spin space. \end{abstract}

\maketitle\section{Introduction}
Turbulence in classical fluids has been studied for a long time. It is an important theme in modern physics, but is not yet understood sufficiently \cite{Frisch}. Great progress in the study of turbulence was made by Kolmogorov \cite{K41}, who suggested that the spectrum of the kinetic energy exhibits a $-5/3$ power law in fully developed isotropic homogeneous turbulence. This is called the Kolmogorov $-5/3$ power law. 

Recently, quantum turbulence (QT) in one-component atomic Bose--Einstein condensates (BECs) has been investigated. Some theoretical and numerical studies show that the Kolmogorov $-5/3$ power law appears in QT \cite{MK-07}. QT in an atomic BEC has been experimentally realized by Henn $\it{et al.}$  \cite{Henn09}, but the Kolmogorov $-5/3$ power law has not yet been observed. 

Multi-component atomic BECs can be realized \cite{Kasamatsu05}, described by more than two macroscopic wave functions, which means that the degrees of freedom in multi-component BECs is larger than in one-component BECs. Thus, various behaviors that do not appear in one-component BEC are expected to occur in this system. The hydrodynamical instability in two-component BECs, such as Rayleigh--Taylor, Kelvin--Helmholtz and Richtmyer--Meshkov instability, has been studied using the Gross--Pitaevskii (GP) equation \cite{Sasaki09, Takeuchi10, Bezett10}. The turbulence in a two-component BEC has been numerically and theoretically investigated, and a tangle of two kinds of quantized vortices was found to form in the system \cite{Ishino10, Ishino11}. 

Another multi-component BEC system is the spinor BEC, comprised of condensates corresponding to different hyperfine states of an atom. We investigated the turbulence in a spinor BEC, namely spin turbulence, in our previous work \cite{Fujimoto11}. In the prior research, we generated the spin turbulence by a counterflow instability, finding that, in the ferromagnetic case, the spectrum of the spin-dependent interaction energy exhibits a $-7/3$ power law. This power law can be understood by scaling analysis with some assumptions. One assumption is that the total density of the condensate is uniform. However, in the actual experiments, the condensates are trapped by a harmonic potential that is not uniform. Therefore, whether the $-7/3$ power law appears in the trapped system is an important issue.

 In this paper, we report the numerical observation of the $-7/3$ power law in a trapped spin-1 spinor BEC. Our study is based on a numerical calculation of the GP equation. The spin turbulence in the trapped system arises from the instability of the helical structure of the spin density vector in the initial state. This helical structure has been experimentally realized by Vengalattore $\it{et al}$. \cite{Vengalattore08}.  We find that, in the ferromagnetic case, the $-7/3$ power law also appears in the trapped system if the system size is much larger than the spin coherence length. This paper is organized as follows. Section II describes the formulation. The results of the numerical simulation of the GP equation are described in Sec. III. Here we show how the helical structure of the spin density vector becomes unstable to the development of spin turbulence and the $-7/3$ power law appears in the spin-dependent interaction energy. In Sec. IV, in order to understand the relation between the energy spectra and the motion of the spin density vector, we discuss the characteristic time of the spin, the orbits in the spin space and the size of the BEC cloud for sustaining the power law. Section V is devoted to conclusions. 
 
 \section{Formulation}
 \subsection{Gross--Pitaevskii equation}
 We consider a trapped spin-1 spinor BEC at zero temperature, which is well described by the macroscopic wave functions $\psi _m$ ($m = 1$, 0, $-1$). Then, the wave functions $\psi _m$ obey the GP equation \cite{Ohmi98, Ho98}:\begin{equation}i\hbar \frac{\partial}{\partial t} \psi _{m} =  (-\frac{\hbar ^2 }{2M} \nabla ^2 + V) \psi _{m} + c_{0} n \psi _{m} + c_{1} \sum _{n=-1} ^{1} \bm{s} \cdot \bm{S} _{mn} \psi _{n}, \label{3dGP}\end{equation}where $M$ and $V$ are the mass of a particle and the trapping potential. In our study, the potential is given by $V = M \omega ^{2}(x^{2}+y^{2})/2 + M \omega _{z}^{2} z^{2}/2$. The total density $n$ and the spin density vector $s_{i}$ ($i = x, y, z$ ) are given by $n =  \sum _{m=-1} ^{1}|\psi _m|^2$ and  $s_{i} = \sum _{m,n = -1}^{1} \psi _{m}^{*} (S_{i})_{mn} \psi _{n}$, where $(S_{i})_{mn}$ are the spin-1 matrices.  The parameters $c_{0}$ and $c_{1}$ are the coefficients of the spin-independent and dependent interactions, which are expressed by $4 \pi \hbar ^{2}(a_{0} + 2a_{2})/3M$ and $4 \pi \hbar ^{2}(a_{2} - a_{0})/3M$ with $s$ wave scattering lengths $a_{0}$ and $a_{2}$. 
 
 The total energy $E$ is given by \begin{eqnarray}\nonumber E = \int \sum _{m=-1} ^{1} [\psi ^{*} _{m} (-\frac{\hbar ^{2}}{2M}\nabla ^2 + V) \psi _{m}] d \bm{r} \\ +\frac{c_{0}}{2} \int n^{2} d \bm{r} + \frac{c_{1}}{2} \int \bm{s} ^{2} d \bm{r}. \label{total energy}\end{eqnarray}The spin-dependent interaction energy is the last term with the coefficient $c_{1}$ of the right hand side of Eq. (\ref{total energy}), which is characteristic of the spinor BEC. The ground state in a uniform system without a magnetic field is ferromagnetic for $c_{1}<0$ and polar for $c_{1}>0$. This interaction exchanges the particles between the different condensates, so the particle number in each component is not conserved.  Thus, the instability and turbulence in spinor BECs are expected to have properties that one- and two-component BECs do not. 
 
 We consider that the condensate is strongly confined in the $z$ direction($\omega \ll \omega _{z}$). In this case, we can approximately separate the degrees of freedom of the macroscopic wavefunctions as $\psi _{m} (x,y,z,t) = \bar{\psi} _{m} (x,y,t) f(z)$. We assume $f(z) = (1/2 \pi a_{hz}^{2})^{1/4}{\rm{e}}^{-z^{2}/4a_{hz}^{2}}$ with $a_{hz} = (\hbar /2M\omega _{z})^{1/2}$. Then, the GP equation (\ref{3dGP}) is reduced to \begin{equation}i\hbar \frac{\partial}{\partial t} \bar{\psi} _{m} =  (-\frac{\hbar ^2 }{2M} \bar{\nabla} ^2 + \bar{V}) \bar{\psi} _{m} + \bar{c}_{0} \bar{n} \bar{\psi} _{m} + \bar{c}_{1} \sum _{n=-1} ^{1} \bar{\bm{s}} \cdot \bm{S} _{mn} \bar{\psi} _{n} \label{2dGP}\end{equation}with $\bar{\nabla}^{2} = \partial ^{2} / \partial x ^{2} + \partial ^{2} / \partial y ^{2}$, $\bar{V} = M \omega ^{2}(x^{2}+y^{2})/2 $, $\bar{c}_{0} = c_{0}/2\sqrt{\pi}a_{hz}$, $\bar{c}_{1} = c_{1}/2\sqrt{\pi}a_{hz}$, $\bar{n} =  \sum _{m=-1} ^{1}|\bar{\psi} _m|^2$ and $\bar{s}_{i} = \sum _{m,n = -1}^{1} \bar{\psi} _{m}^{*} (S_{i})_{mn} \bar{\psi} _{n}$. We consider experiments with $^{87}$Rb, which exhibits a ferromagnetic interaction. Thus, we use $M = 1.42 \times 10^{-25}$ kg, $a_{0} = 5.39 \times 10^{-9}$ m, $a_{2} = 5.31 \times 10^{-9}$ m, $N = 3 \times 10^{5}$, $\omega = 2 \pi \times 20$ /s and $\omega _{z}= 2 \pi \times 600$ /s. 
 
 \subsection{Numerical method}We use the Crank--Nicholson method to numerically calculate the GP equation (\ref{2dGP}). The coordinate is normalized by the length $a_{h} = (\hbar /2M \omega)^{1/2} \sim 1.72 \mu$m and the box size is $80a_{h} \times 80a_{h}$. Space in the $x$ and $y$ directions is discretized into $1024 \times 1024$ bins. The time is normalized by the frequency $\omega$ of the trapping potential in the $x$ and $y$ directions.\begin{figure}[t]\begin{center}\includegraphics[keepaspectratio, width=9cm,clip]{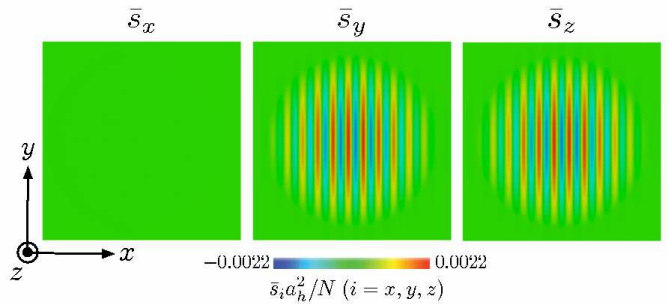} \caption{(Color online) Profiles of each component of the spin density vector in the initial state. The box size is $40 a_{h} \times 40 a_{h}$, where $a_{h}$ is $(\hbar /2M \omega)^{1/2} \sim 1.72 \mu$m.} \label{fig:initial_state}\end{center}\end{figure}
 
 \subsection{Initial state}We use an initial state with a helical structure of the spin density vector to obtain the spin turbulence in the trapped spin-1 spinor BEC. In the following, we show the mathematical expression of this state. The macroscopic wave functions $\bar{\bm{\psi}} = (\bar{\phi}, 0, 0)$ has spin density vector $\bar{\bm{s}} = |\bar{\phi}|^{2} \hat{\bm{e}}_{z}$, where $\hat{\bm{e}}_{j} (j = x, y, z)$ is the unit vector in the $j$ direction. Here, $\bar{\phi}$ can be obtained by the imaginary time step of Eq. (\ref{2dGP}). By multiplying this wave function by the rotation matrix $\hat{U} = e^{-i\alpha \hat{S}_{z}} e^{-i\beta \hat{S}_{y}} e^{-i\gamma \hat{S}_{z}}$ in the spin space, we obtain\begin{equation}\begin{pmatrix} \bar{\psi} _{1} \\\bar{\psi} _{0} \\\bar{\psi} _{-1}\end{pmatrix}= \bar{\phi} e^{-i\gamma}\begin{pmatrix} e^{-i\alpha} {\rm{cos}}^{2}\frac{\beta}{2}  \\\frac{1}{\sqrt{2}} {\rm{sin}} \beta  \\e^{i\alpha} {\rm{sin}}^{2}\frac{\beta}{2}\end{pmatrix}, \label{initial state1}\end{equation}where $\alpha$, $\beta$, $\gamma$ are the Euler angles. Then, the spin density vector is expressed by $\bar{\bm{s}} = |\bar{\phi}|^{2}({\rm{sin}}\beta \hspace{0.8mm}{\rm{cos}}\alpha \hspace{0.8mm} \hat{\bm{e}}_{x} + {\rm{sin}}\beta \hspace{0.8mm}  {\rm{sin}}\alpha \hspace{0.8mm} \hat{\bm{e}}_{y} + {\rm{cos}}\beta \hspace{0.8mm} \hat{\bm{e}}_{z})$. Therefore, using $\alpha = \pi /2$, $\beta = k_{h}x$ and $\gamma = 0$, the spin density vector has a helical structure with wave number $k_{h}$. In this paper, we consider the case of $k_{h} \sim 1.22$ /$\mu$m. Thus, Eq. (\ref{initial state1}) becomes\begin{equation}\begin{pmatrix} \bar{\psi} _{1} \\\bar{\psi} _{0} \\\bar{\psi} _{-1}\end{pmatrix}= \bar{\phi}\begin{pmatrix} -i \hspace{0.8mm} {\rm{cos}}^{2}(k_{h}x/2) \\\frac{1}{\sqrt{2}} {\rm{sin}}(k_{h}x) \\i \hspace{0.8mm} {\rm{sin}}^{2}(k_{h}x/2)\end{pmatrix}.\label{initial state2}\end{equation}Figure \ref{fig:initial_state} shows the profile of each component of the spin density vector in the initial state of Eq.(\ref{initial state2}). The helical structure has been experimentally realized by Vengalattore $\it{et al.}$  \cite{Vengalattore08}, where they prepared the structure by means of a magnetic field, investigating how the structure becomes unstable to changes into some disordered state through observing the spin density vector. In our calculations, we add some small white noise to the initial state of Eq. (\ref{initial state2}). The noise causes the particle number of each component to fluctuate by $0.1$--$0.3 \%$. This is consistent with experimental results \cite{Sadler06}. 
 
 \subsection{Spectrum of spin-dependent interaction energy}We derive an expression for the spectrum of the spin-dependent interaction energy. The spin-dependent interaction energy $E_{s}$ is given by \begin{equation}E_{s} = \frac{c_{1}}{2} \int \bar{\bm{s}}(\bar{\bm{r}})^{2} d\bar{\bm{r}}\end{equation}with $ \bar{\bm{r}}=(x,y)$.  We expand the spin density vector $\bar{\bm{s}}(\bar{\bm{r}})$ with plane waves:$\bar{\bm{s}}(\bar{\bm{r}}) = \sum _{ \bar{\bm{k}}} \tilde{\bm{s}}(\bar{\bm{k}}) e^{i \bar{\bm{k}}\cdot \bar{\bm{r}}} $with $ \bar{\bm{k}}=(k_{x},k_{y})$. Then, the spin-dependent interaction energy $E_{s}$ is represented by $\tilde{\bm{s}}(\bar{\bm{k}})$ as\begin{equation}E_{s} = \frac{c_{1}A}{2} \sum _{\bar{\bm{k}}} |\tilde{\bm{s}}(\bar{\bm{k}})|^{2},  \end{equation}where $A$ is the area of the system. Therefore, the energy spectrum of the spin-dependent interaction energy is given by\begin{equation}E_{s} (k) = \frac{c_{1}A}{2 \Delta k} \sum _{k<|\bar{\bm{k}}_{1}|<k+\Delta k} |\tilde{\bm{s}}(\bar{\bm{k}}_{1})|^{2},  \label{spectrum_of_spin}\end{equation}where $\Delta k$ is $2\pi /L$ for a system size $L$.\begin{figure}[t]\begin{center}\includegraphics[keepaspectratio, width=9cm,clip]{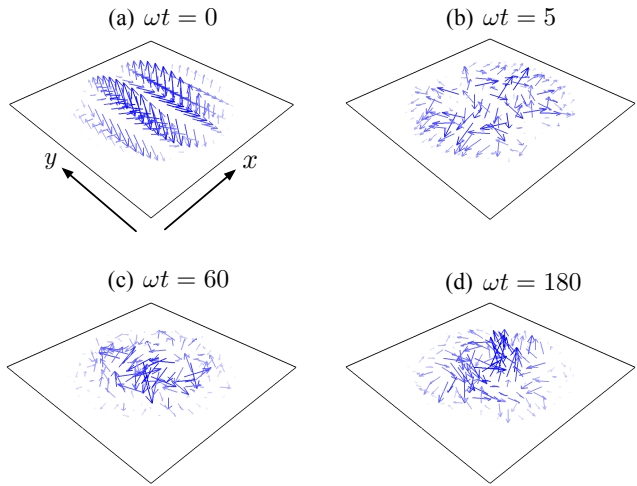} \caption{(Color online) Profiles of the spin density vector at $\omega t = 0$, 5, 60, 180. The shading of the arrows denotes the amplitude of the spin density vector. The box size is $40 a_{h} \times 40 a_{h}$. } \label{fig:spin density vector}\end{center}\end{figure}
 
 \section{Numerical results}
 We show our numerical results with the GP equation (\ref{2dGP}) starting from the initial state of Eq. (\ref{initial state2}) \cite{movie}. Our calculation finds that the helical structure in the trapped system leads to spin turbulence and the spectrum of the spin-dependent interaction energy exhibits the $-7/3$ power law. 
 
The helical structure of the spin density vector in the trapped system is unstable, generating the spin turbulence. Figure \ref{fig:spin density vector} (a) shows the profile of the spin density vector in the initial state. This helical structure immediately collapses as shown in Fig. \ref{fig:spin density vector} (b). As time goes by, the spin density vector is significantly disturbed, pointing in various directions, as shown in Figs. \ref{fig:spin density vector} (c) and (d). Thus, spin turbulence in the trapped system is realized. \begin{figure*}[t]\begin{center}\includegraphics[keepaspectratio, width=12cm,clip]{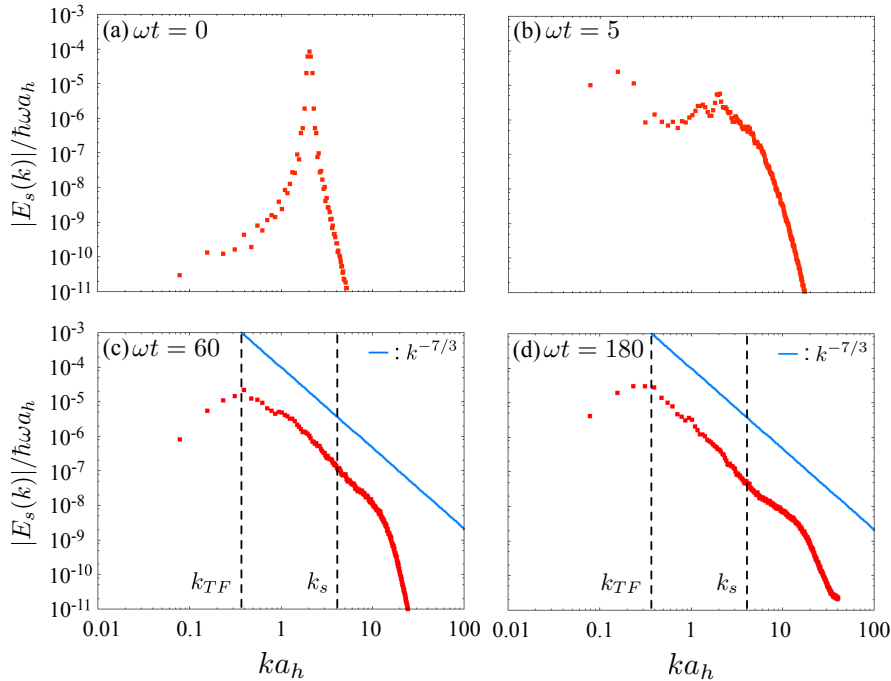} \caption{(Color online) Spectrum of the spin-dependent interaction energy at $\omega t = 0$, 5, 60, 180. $k_{s}$ and $k_{TF}$ are the wave numbers corresponding to the spin coherence length $\xi _{s} = \hbar/\sqrt{2M|c_{1}|n_{0}}$ and the Thomas--Fermi radius $R_{TF}$, where $n_{0}$ is the total density at the center in the initial state. Their expressions are $k_{s} = 2 \pi /\xi _{s}$ and $k_{TF} = 2 \pi /R_{TF}$.}  \label{fig:spectrum}\end{center}\end{figure*}

The spectrum of the spin-dependent interaction energy exhibits the $-7/3$ power law in the spin turbulence in the trapped system. Figures \ref{fig:spectrum} show the time development of the spectrum, which is numerically calculated by Eq. (\ref{spectrum_of_spin}). In the initial state, the spectrum in Fig. \ref{fig:spectrum} (a) has a sharp peak at $ka_{h} \sim 2$ corresponding to the wave number $k_{h}$ of the helical structure. As the instability develops, modulation of the spin density vector with various wave numbers is excited. Thus, as shown in Fig. \ref{fig:spectrum} (b), the spectrum loses the sharp peak, letting the energy flow into both the low and high wave number regions. As time passes, the spin density vector is significantly disturbed and the system exhibits spin turbulence. Then, the spectrum develops to exhibit the $-7/3$ power law in the wave number region $k_{TF} < k < k_{s}$, as shown in Fig. \ref{fig:spectrum} (c). Here $k_{TF}$ and $k_{s}$ are the wave numbers corresponding to the Thomas--Fermi radius $R_{TF}$ and the spin coherence length $\xi _{s} = \hbar/\sqrt{2M|c_{1}|n_{0}}$. This $-7/3$ power law in the spectrum continues at least until $\omega t = 180$. Figure \ref{fig:spectrum} (d) shows the spectrum at $\omega t = 180$.

\section{Discussion}
\subsection{Motion of the spin density vector and the $-7/3$ power law}
The appearance of the $-7/3$ power law is accompanied by characteristic behavior of the spin density vector. We should note the time scale of the motion as well as the randomness of the spin density vector. After the initial helical structure becomes unstable, the spin vectors exhibit relatively large-scale, slow motion. The helical structure then disappears and the vectors become disordered. As the $-7/3$ power law appears, the vectors begin to perform rapid, fine oscillations around random directions; the motion of each spin vector appears to be frozen to a random direction.  

This can be understood qualitatively by the following discussion. In our previous study \cite{Fujimoto11}, we performed Kolmogorov-type scaling analysis for the equation of motion of the spin density vector and thus obtained the $-7/3$ power law. In the wave number region sustaining the $-7/3$ power law, the characteristic time $t_{s}(k)$ of the mode of wave number $k$ is expressed by \begin{equation}t_{s}(k) \sim \epsilon ^{-1/3} k^{-3/4}, \label{time_k}\end{equation}where $\epsilon$ is the energy flux flowing from low to high wave numbers. Equation (\ref{time_k}) is obtained from $\epsilon \sim k^{-4}t_{s}^{-3}$, which means that the energy flux is independent of the wave number \cite{Fujimoto11}. Equation (\ref{time_k}) shows that the characteristic time $t_{s}$ shortens when the wave number is high. 

As the spin density vectors become disordered and the high $k$ modes (small scale) are excited, the characteristic time scale becomes short, leading to the rapid, fine oscillations. On the other hand, the motion with low $k$ modes (large scale) has a long characteristic time $t_{s}$. Thus, the rough direction of the spin density vector at each position is largely fixed and changes only slowly. 

Thus the motion of the spin density vector is closely related to the advent of the $-7/3$ power law. \begin{figure*}[t]\begin{center}\includegraphics[keepaspectratio, width=12cm,clip]{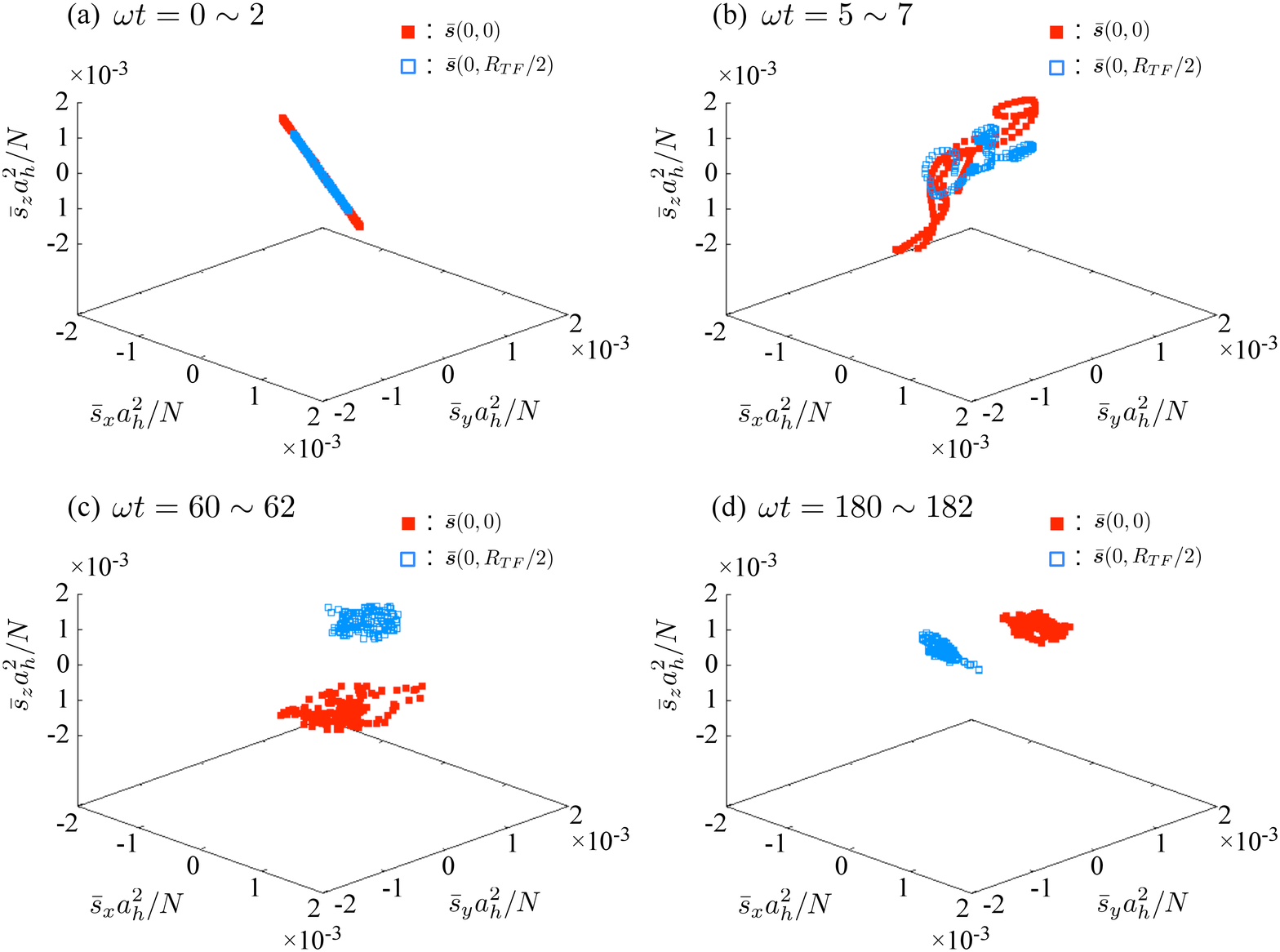} \caption{(Color online) Orbits of the spin density vector at two positions $(0,0)$ (red filled squares) and $(0, R_{TF}/2)$ (blue empty squares) inside the BEC cloud for the periods (a) $\omega t = 0$--2, (b) 5--7, (c) 60--62 and (d) 180--182. The spin density vector is normalized by $N/a_h^2$.}   \label{fig:orbits}\end{center}\end{figure*}\begin{figure}[t]\begin{center}\includegraphics[keepaspectratio, width=9cm,clip]{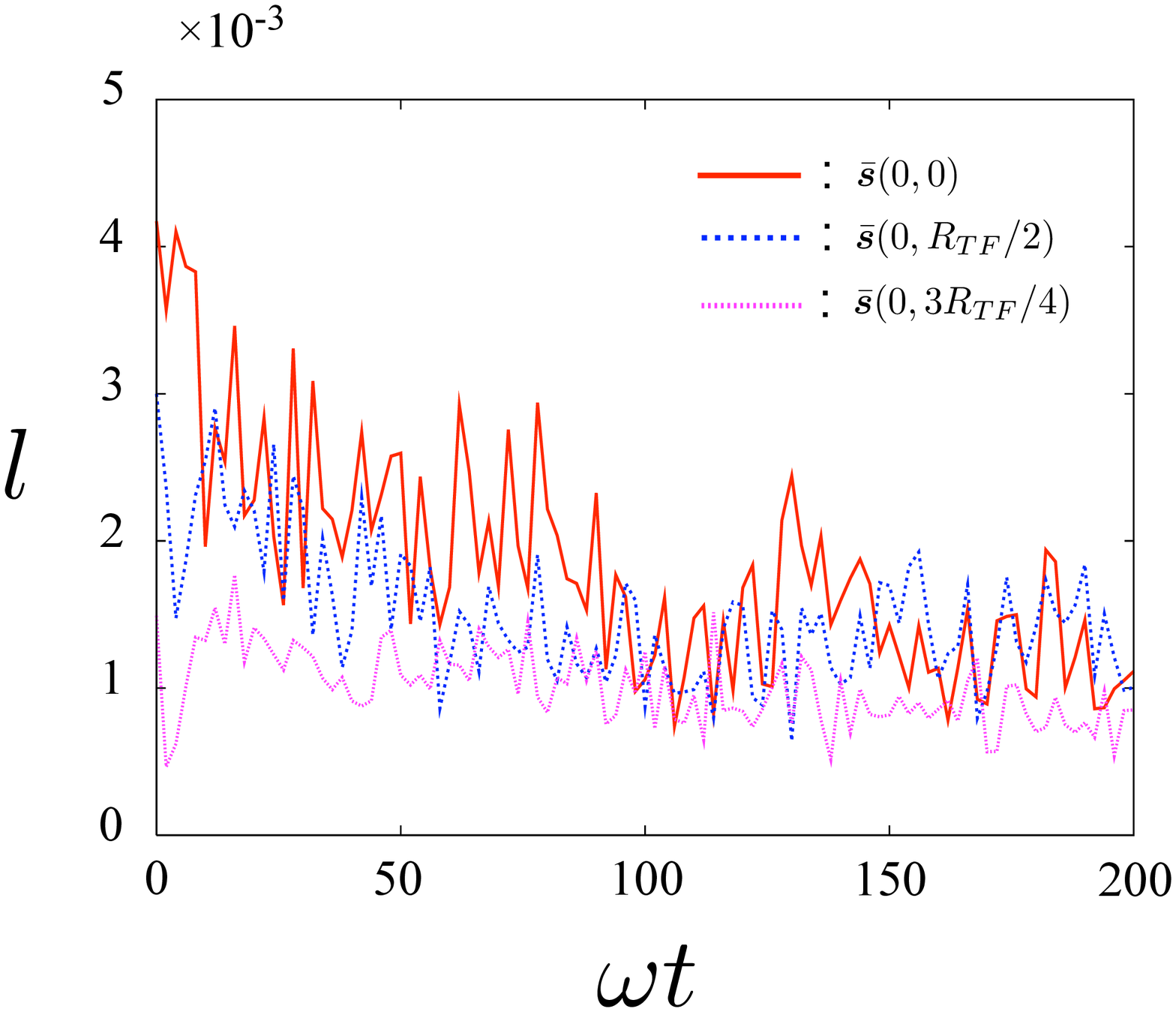} \caption{(Color online) Time development of the size of the orbits of the spin density vector in the spin space of Fig. 4. The three curves correspond to the three positions $(0,0)$, $(0,R_{TF}/2)$ and $(0, 3R_{TF}/4)$ inside the BEC cloud. The size $l$ at time $\tau=\omega t$ refers to the maximum distance between two points in an orbit during $\tau$ to $\tau +2$.}  \label{fig:size}\end{center}\end{figure}

\subsection{Orbits in spin space}
In order to reveal these properties, it is useful to consider the orbits of the dynamical variables in phase space, which is an approach often used in the research field of nonlinear dynamics. Figure \ref{fig:orbits} shows the orbits of the normalized spin density vectors in the $(\bar{s}_{x}a_h^2/N,\bar{s}_{y}a_h^2/N,\bar{s}_{z}a_h^2/N)$ space for different time stages. The two orbits correspond to the spin density vector at the positions $(0,0)$ and $(0, R_{TF}/2)$ inside the BEC cloud. The amplitude of the spin density vector depends on the position inside the BEC cloud and is maximum at the center; such a dependence of the amplitude is reflected in the orbits of the  spin at different positions. We discuss the time development of the orbits of Fig. 4 with reference to Fig. 2 and Fig. 3.  The periodic orbits in the early stage (Fig. 4 (a)) indicate some coherent motion of the initial helical structure. Some of the orbits become unstable (Fig. 4 (b)) after the helical structure become unstable. Then, the spin density vectors start to become disordered (Fig. 2 (b)) and the energy spectrum becomes broad but does not yet exhibit the $-$7/3 power law (Fig. 3(b)). After a time, the two orbits become separated, as shown in Fig. 4(c), and independently localized in some places in the spin space, generating rapid, fine oscillation in those areas. During this process the spin density vectors become disordered and frozen (Fig. 2(c)) and the energy spectrum exhibits the $-$7/3 power law (Fig. 3(c)). This behavior continues until $\omega t = 180$, while the orbits appear to shrink from Fig. 4 (b) through (c) to (d). After the power law is established, the orbits move very slowly in the spin space, as shown from Fig. 4 (c) to (d).      

In order to measure the shrinkage quantitatively, we introduce the size $l$ of the orbit in the spin space, where the size $l$ at time $\tau=\omega t$ is defined by the maximum distance between two points in an orbit during $\tau$ to $\tau +2$.    Figure \ref{fig:size} shows the sizes $l(\tau)$ of orbits at three different positions $(0,0)$, $(0, R_{TF}/2)$ and $(0, 3R_{TF}/4)$ inside the BEC cloud. The sizes of the orbits are found to fluctuate and shrink  as time passes; the size is generally smaller at outer positions in the cloud, presumably because of the smaller amplitude of the spin density vector. 

It is interesting that the freezing of the separated spin orbits can determine the power law of the energy spectrum.   

 \subsection{Size of condensate and the $-7/3$ power law} We discuss the relation between the size of the condensate and the $-7/3$ power law. We confirm that if the Thomas--Fermi radius $R_{TF}$ is not sufficiently larger than the spin coherence length $ \xi _{s}$, the spectrum of the spin-dependent interaction energy does not exhibit the $-7/3$ power law. This is confirmed by a numerical calculation with $R_{TF}/ \xi _{s} \sim 5$. For the calculations in this paper, the condition $R_{TF}/ \xi _{s} \sim 10$ is satisfied. Therefore, in the experiments, the condition $R_{TF}/ \xi _{s} \geq 10$ could be necessary for observing the $-7/3$ power law.  
 
 \section{Conclusions}
 We study spin turbulence in a trapped spin-1 spinor BEC by numerical calculation of the GP equation (\ref{2dGP}). The spin turbulence in the trapped system is generated by instability of the helical structure of the spin density vector. Our numerical calculation finds that the spectrum of the spin-dependent interaction energy exhibits the $-7/3$ power law, which is consistent with our previous result for a uniform system \cite{Fujimoto11}. We discuss the relation between the $-7/3$ power law and the motion of the spin density vector. 
 
We believe that the $-7/3$ power law can be experimentally observed because of the following reasons. As discussed in Sec. IV C, the spin turbulence in a large trapped system ($R_{TF}/ \xi _{s} \sim 10$) sustains this power law for a long time. All the components of the spin density vector have been experimentally observed by a phase contrast imaging method \cite{Vengalattore08,Sadler06}. The expressions for the spectrum of the spin-dependent interaction energy show that it is possible to obtain a spectrum if the spin density vector is observed everywhere. This is because Eq. (\ref{spectrum_of_spin}) contains only the Fourier component of the vector $\tilde{\bm{s}}(\bm{k})$. Therefore, it is possible to observe this power law. \section*{ACKNOWLEDGMENT}M. T. acknowledges the support of a Grant-in-Aid for Science Research from JSPS (Grant No. 21340104).\end{document}